# Bingham Fluid Flow through Oscillatory Porous Plate with Ion-Slip and Hall Current

Md. Tusher Mollah [1, a)], Muhammad Minarul Islam[1, b)], Mohammad Ferdows[2, c)] and Md. Mahmud Alam[3, d)]

[1] *Department of Mathematics, Bangabandhu Sheikh Mujibur Rahman Science and Technology University Gopalganj-8100, Bangladesh*
[2] *Department of Applied Mathematics, University of Dhaka, Dhaka-1000, Bangladesh*
[3] *Mathematics Discipline, Khulna University, Khulna-9208, Bangladesh*

a) Corresponding author: tusher.bsmrstu@gmail.com
b) minarul_math@yahoo.com
c) ferdows@du.ac.bd
d) alam_mahmud2000@yahoo.com

**Abstract.** The numerical approach has been performed to study the Bingham fluid flow through an oscillatory porous plate with Ion-Slip and Hall current. Initially, at time; $t = 0$ both the fluid and the upper plate are at rest. At time; $t > 0$ the upper plate begins to oscillate in its own plane while the lower plate is stationary. The lower plate temperature is constant while the upper plate temperature has oscillated. A uniform magnetic field is applied perpendicular to the plates. To obtain the dimensionless equations from the governing non-linear partial differential equations, the usual transformations have been used. The explicit finite difference technique has been applied to solve the obtained dimensionless equations. The MATLAB R2015a has been used for numerical simulation. For the accuracy of the numerical technique, the stability and convergence criteria have been discussed and the system has found to be converged for $P_r \geq 0.08$, $\beta_i \geq 2$, $H_a \leq 20$, $k_0 \leq 8\,(k \neq 2)$ and $R_e \geq -0.011$ with $\beta_e = 0.10$, $E_c = 0.10$, $\Delta Y = 0.05$ and $\Delta \tau = 0.0001$. The steady-state solution has achieved at the dimensionless time $\tau = 2.00$. At the steady-state time, the effect of several parameters on the flow patterns, local shear stress and the Nusselt number have been shown graphically.

## INTRODUCTION

The Magnetohydrodynamic (MHD) fluid flow through a porous medium have considerable attention in various scientific and engineering applications. Ample applications of such flows are encountered in the process of petroleum, irrigation problems, filtration and purification processes, biological systems, as well as paper and polymer composite industries. Numerous investigations have been carried out on various aspects of MHD flows of non-Newtonian fluid passing through porous plate oscillating in its own plane.

In this regard, the first exact solution of Navier-Stokes' equation has been studied by Stokes [1] which is concerned with the flow of viscous incompressible fluid past a horizontal plate oscillating in its own plane. Datta and Jana [2] considered the oscillatory MHD flow past a flat plate with Hall effects. The MHD oscillatory flow on free convection-radiation through a porous medium with constant suction velocity has been investigated by El-Hakiem [3]. The unsteady MHD Hartmann–Couette flow and heat transfer in a Darcian channel with Hall current, Ion-slip, viscous and Joule heating effects has been considered by Bég el al. [4]. The finite element solution of heat and mass transfer in the MHD flow of a viscous fluid past a vertical plate under oscillatory suction velocity has been studied by Rao et al. [5]. The MHD free convection and mass transfer flow through a vertical oscillatory porous plate with Hall, Ion-slip currents and heat source in a rotating system has been investigated by Hossain et al. [6]. The impact silver nanoparticles on MHD free convection flow of Jeffrey fluid over an oscillating vertical plate embedded in a porous medium has been studied by Zin et al. [7]. The MHD viscoelastic fluid flow along an infinite oscillating porous plate with the heat source and thermal diffusion has been considered by Sharmin and Alam [8].

Along with these studies our aim is to study the Bingham fluid flow through the oscillatory porous plate with Ion-Slip and Hall current. The present study is concerned with the generalized Ohm's law in the presence of Ion-Slip and Hall current. Also, the upper plate oscillation and the temperature oscillation are considered. The viscous dissipation, pressure gradient, porous medium, uniform magnetic field and the rheology of Bingham fluid have been involved in the governing equations. The explicit finite difference technique has been used to solve the dimensionless governing equations. The steady-state solutions and the effect of interesting parameters have been shown graphically.

## MATHEMATICAL FORMULATION

Consider the laminar and incompressible non-Newtonian Bingham fluid, which is flowing between two infinite horizontal plates, located at $y = \pm h$ planes and extend from $x = 0$ to $\infty$ and from $z=0$ to $\infty$. Initially, at time; $t = 0$ both the fluid and the upper plate are at rest and their temperature is $T_1$. At time; $t > 0$ the upper plate begins to oscillate in its plane with the velocity $U_0 \cos \omega t$ in a given direction, also the plate temperature is raised to $T_2$ and a periodic temperature is assumed to be superimposed on this mean constant temperature of the plate. The lower plate is stationary and its temperature is $T_1$, where $T_2 > T_1$. A constant pressure gradient $\frac{dp}{dx}$ is applied on the fluid along the X-direction and a uniform magnetic field $B_0$ is acted perpendicular to the X-direction, which is undisturbed whereas a very small magnetic Reynolds number. A Z-component of the velocity is expected to arise due to consideration of Hall current. Thus the fluid velocity vector is given as $\mathbf{q} = u\mathbf{i} + v\mathbf{j} + w\mathbf{k}$.

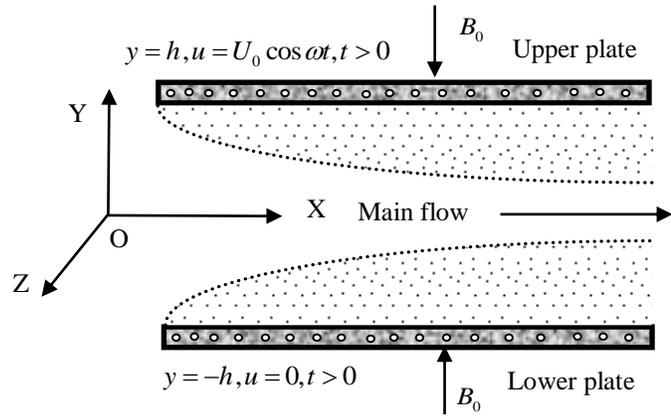

**Figure 1**. The physical configuration of the problem

To obtain the dimensionless governing equations, the dimensionless quantities can be defined as follows:
$$X = \frac{x}{h}, Y = \frac{y}{h}, U = \frac{u}{U_0}, V = \frac{v}{U_0}, W = \frac{w}{U_0}, P = \frac{p}{\rho U_0^2}, \tau = \frac{tU_0}{h}, \theta = \frac{T - T_1}{T_2 - T_1}, \bar{\mu} = \frac{\mu}{K}, \eta = \frac{\omega h}{U_0}$$

The obtained dimensionless governing non-linear coupled partial differential equations are given as follows:

$$\frac{\partial U}{\partial X} + \frac{\partial V}{\partial Y} = 0 \tag{1}$$

$$\frac{\partial U}{\partial \tau} + U\frac{\partial U}{\partial X} + V\frac{\partial U}{\partial Y} = -\frac{dP}{dX} + \frac{1}{R_e}\left[\frac{\partial}{\partial Y}\left(\bar{\mu}\frac{\partial U}{\partial Y}\right) - \frac{H_a^2}{\alpha_e^2 + \beta_e^2}(\alpha_e U + \beta_e W)\right] - k_0 U \tag{2}$$

$$\frac{\partial W}{\partial \tau} + U\frac{\partial W}{\partial X} + V\frac{\partial W}{\partial Y} = \frac{1}{R_e}\left[\frac{\partial}{\partial Y}\left(\bar{\mu}\frac{\partial W}{\partial Y}\right) - \frac{H_a^2}{\alpha_e^2 + \beta_e^2}(\alpha_e W - \beta_e U)\right] - k_0 W \tag{3}$$

$$\frac{\partial \theta}{\partial \tau} + U\frac{\partial \theta}{\partial X} + V\frac{\partial \theta}{\partial Y} = \frac{1}{P_r}\frac{\partial^2 \theta}{\partial Y^2} + E_c\bar{\mu}\left[\left(\frac{\partial U}{\partial Y}\right)^2 + \left(\frac{\partial W}{\partial Y}\right)^2\right] + \frac{E_c H_a^2}{\alpha_e^2 + \beta_e^2}\cdot\left[U^2 + W^2\right] \tag{4}$$

where, $\bar{\mu} = 1 + \dfrac{\tau_D}{\sqrt{\left(\dfrac{\partial U}{\partial Y}\right)^2 + \left(\dfrac{\partial W}{\partial Y}\right)^2}}$ \tag{5}

and the dimensionless initial and boundary conditions can be written as follows:

$$\tau \leq 0, \quad U = 0, W = 0, \theta = 0 \quad \text{everywhere} \tag{6}$$

$$U = 0, W = 0, \theta = 0 \text{ at } X = 0$$

$$\tau > 0, \quad U = 0, W = 0, \theta = 0 \text{ at } Y = -1$$

$$U = \cos\eta\tau, W = 0, \theta = 1 + \cos\eta\tau \text{ at } Y = 1 \tag{7}$$

The non-dimensional parameters are given as follows:

$\tau_D = \dfrac{\tau_0 h}{K U_0}$ (Bingham number or dimensionless yield stress), $R_e = \dfrac{\rho U_0 h}{K}$ (Reynolds number), $P_r = \dfrac{\rho c_p U_0 h}{k}$ (Prandtl number), $E_C = \dfrac{U_0 K}{\rho c_p h (T_2 - T_1)}$ (Eckert number), $Ha^2 = \dfrac{\sigma B_0^2 h^2}{K}$ (Hartmann number squared) and $k_0 = \dfrac{\upsilon^2}{k U_0^2}$ (Permeability of porous medium).

## SHEAR STRESS AND NUSSELT NUMBER

The effects of various parameters on shear stress have been studied from the velocity profile. The local shear stress in $X$- direction for stationary wall is $\tau_{L1} \equiv \left[\mu\sqrt{\left(\dfrac{\partial U}{\partial Y}\right)^2 + \left(\dfrac{\partial W}{\partial Y}\right)^2}\right]_{Y=-1}$ and for oscillatory wall is $\tau_{L2} \equiv \left[\mu\sqrt{\left(\dfrac{\partial U}{\partial Y}\right)^2 + \left(\dfrac{\partial W}{\partial Y}\right)^2}\right]_{Y=1}$. Also the effects of various parameters on Nusselt number have been studied from the temperature profile. The local Nusselt number in $X$- direction for stationary wall is $Nu_{L1} \equiv \dfrac{\left(\dfrac{\partial T}{\partial Y}\right)_{Y=-1}}{-T_m}$ and for oscillatory wall is $Nu_{L2} \equiv \dfrac{\left(\dfrac{\partial T}{\partial Y}\right)_{Y=1}}{-(T_m - 1)}$, where $T_m$ is the dimensionless mean fluid temperature and is given by $T_m = \dfrac{\int_{-1}^{1} U\theta \, dY}{\int_{-1}^{1} U \, dY}$.

## NUMERICAL TECHNIQUE

To solve the dimensionless coupled partial differential equations (1) to (5) by the explicit finite difference method subject to the boundary conditions, a set of finite difference equations is required. For which, the region within the boundary layer is divided into a grid or mesh of lines parallel to $X$ and $Y$ axes where $X$-axis is taken along the plate and $Y$-axis is normal to the plate as shown in Fig. 1.

Here it is considered that the height of the plate is $X_{max}\,(= 40)$ i.e. $X$ varies from 0 to 40 and width of the plate is $Y_{max}\,(= 2)$ as corresponding to $Y \to \infty$ i.e. $Y$ varies from 0 to 2. Also, the appropriate grid spacing $(m, n) = (40, 40)$ are chosen in the $(X, Y)$ directions. The constant mesh sizes $\Delta X$ and $\Delta Y$ are assumed along $X$ and $Y$ directions respectively, where, $\Delta X = 1.0\,(0 \leq x \leq 40)$, $\Delta Y = 0.05\,(0 \leq y \leq 2)$ with the smaller time-step, $\Delta \tau = 0.0001$.

Let $U'$, $W'$ and $\theta'$ denote the values of $U$, $W$ and $\theta$ at the end of a time-step respectively. Using the explicit finite difference approximation the appropriate set of finite difference equations are obtained as follows:

$$\frac{U_{i,j} - U_{i-1,j}}{\Delta X} + \frac{V_{i,j} - V_{i,j-1}}{\Delta Y} = 0 \tag{8}$$

$$\frac{U'_{i,j} - U_{i,j}}{\Delta \tau} + U_{i,j}\frac{U_{i,j} - U_{i-1,j}}{\Delta X} + V_{i,j}\frac{U_{i,j} - U_{i,j-1}}{\Delta Y} = -\frac{dP}{dX} - k_0 U_{i,j}$$
$$+ \frac{1}{R_e}\left[\left(\frac{\bar{\mu}_{i,j} - \bar{\mu}_{i,j-1}}{\Delta Y}\right)\left(\frac{U_{i,j} - U_{i,j-1}}{\Delta Y}\right) + \bar{\mu}_{i,j}\left(\frac{U_{i,j+1} - 2U_{i,j} + U_{i,j-1}}{(\Delta Y)^2}\right) - \frac{H_a^2}{\alpha_e^2 + \beta_e^2}\left(\alpha_e U_{i,j} + \beta_e W_{i,j}\right)\right] \tag{9}$$

$$\frac{W'_{i,j}-W_{i,j}}{\Delta\tau}+U_{i,j}\frac{W_{i,j}-W_{i-1,j}}{\Delta X}+V_{i,j}\frac{W_{i,j}-W_{i,j-1}}{\Delta Y}=-k_0 W_{i,j}$$
$$+\frac{1}{R_e}\left[\left(\frac{\bar{\mu}_{i,j}-\bar{\mu}_{i,j-1}}{\Delta Y}\right)\left(\frac{W_{i,j}-W_{i,j-1}}{\Delta Y}\right)+\bar{\mu}_{i,j}\left(\frac{W_{i,j+1}-2W_{i,j}+W_{i,j-1}}{(\Delta Y)^2}\right)-\frac{H_a^2}{\alpha_e^2+\beta_e^2}\left(\alpha_e W_{i,j}-\beta_e U_{i,j}\right)\right] \quad (10)$$

$$\frac{\theta'_{i,j}-\theta_{i,j}}{\Delta\tau}+U_{i,j}\frac{\theta_{i,j}-\theta_{i-1,j}}{\Delta X}+V_{i,j}\frac{\theta_{i,j}-\theta_{i,j-1}}{\Delta Y}=\frac{1}{P_r}\frac{\theta_{i,j+1}-2\theta_{i,j}+\theta_{i,j-1}}{(\Delta Y)^2}$$
$$+E_c\left(\bar{\mu}_{i,j}\right)\left[\left(\frac{U_{i,j}-U_{i,j-1}}{\Delta Y}\right)^2+\left(\frac{W_{i,j}-W_{i,j-1}}{\Delta Y}\right)^2\right]+\frac{E_c H_a^2}{\left(\alpha_e^2+\beta_e^2\right)}\cdot\left[\left(U_{i,j}\right)^2+\left(W_{i,j}\right)^2\right] \quad (11)$$

$$\bar{\mu}_{i,j}=1+\frac{\tau_D}{\sqrt{\left(\dfrac{U_{i,j}-U_{i,j-1}}{\Delta Y}\right)^2+\left(\dfrac{W_{i,j}-W_{i,j-1}}{\Delta Y}\right)^2}} \quad (12)$$

and the boundary conditions with the finite difference scheme are given as follows:
$$U_{i,L}=0,\ W_{i,L}=0,\ \theta_{i,L}=0 \text{ at } L=-1 \text{ and } U_{i,L}=\cos\eta\tau,\ W_{i,L}=0,\ \theta_{i,L}=1+\cos\eta\tau \text{ at } L=1$$

Here the subscripts $i$ and $j$ designate the grid points with $X$ and $Y$ coordinates respectively; where, $0\leq\eta\tau\leq\dfrac{\pi}{2}$.

## STABILITY AND CONVERGENCE ANALYSIS

Since an explicit procedure is being used, the analysis will remain incomplete unless the stability and convergence of the finite difference scheme are discussed. For the constant mesh sizes the stability and convergence criteria finally can be written as follows:

$$\frac{U\Delta\tau}{\Delta X}-\frac{|V|\Delta\tau}{\Delta Y}+\frac{\Delta\tau}{2R_e}\frac{H_a^2}{\alpha_e^2+\beta_e^2}\alpha_e+\frac{k_0}{2}\leq 1 \text{ and } \frac{U\Delta\tau}{\Delta X}-\frac{|V|\Delta\tau}{\Delta Y}+\frac{2\Delta\tau}{P_r(\Delta Y)^2}\leq 1$$

Using $\Delta Y=0.05$, $\Delta\tau=0.0001$ and the initial condition, the above equations gives $P_r\geq 0.08$, $\beta_i\geq 2$, $H_a\leq 20$, $k_0\leq 8\ (k\neq 2)$ and $R_e\geq -0.011$ with $\beta_e=0.10$ and $E_c=0.10$.

## RESULTS AND DISCUSSION

In order to point out the physical condition of the developed mathematical model, the steady state numerical values have been computed for the non-dimensional primary velocity $(U)$, secondary velocity $(W)$ and temperature $(\theta)$ distributions within the boundary layer. In association with the stability and convergence criteria, the steady-state solution has obtained at the dimensionless time $\tau=2.00$. At the steady state time, the effect of Ion-slip parameter $(\beta_i=5.0, 7.0 \text{ and } 9.0)$, where $\alpha_e=1+\beta_i\beta_e$ and phase angle $\left(\eta\tau=\dfrac{\pi}{6},\ \dfrac{\pi}{4} \text{ and } \dfrac{\pi}{3}\right)$ on the primary velocity $(U)$, secondary velocity $(W)$ and temperature $(\theta)$ distributions as well as local shear stress at stationary plate $(\tau_{L1})$, local shear stress at oscillatory plate $(\tau_{L2})$, local Nusselt number at stationary plate $(Nu_{L1})$ and local Nusselt number at oscillatory plate $(Nu_{L2})$ are discussed as shown in Fig. 2(b,c,d) and 3(a,b,c,d); where, Hall parameter $(\beta_e=0.10)$, Hartmann number $(H_a=3.00)$, Reynolds number $(R_e=3.00)$, Eckert number $(E_c=0.10)$, Prandtl number $(P_r=0.2)$, Permeability of porous medium $(k_0=0.5)$ and Bingham number $(\tau_D=0.1)$. For brevity, the effects of other parameters are not shown.

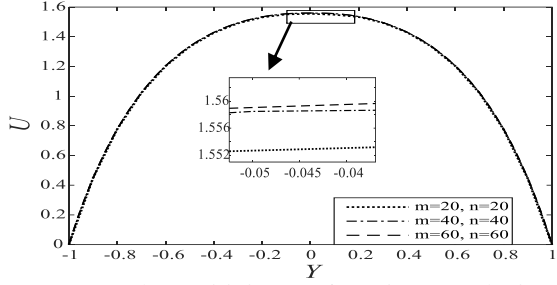
2(a) Mesh Sensitivity test for Primary Velocity

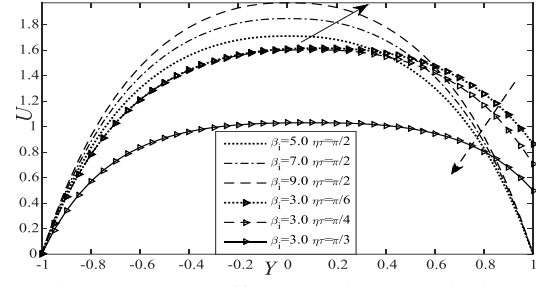
2(b) Parameters Effect on Primary Velocity

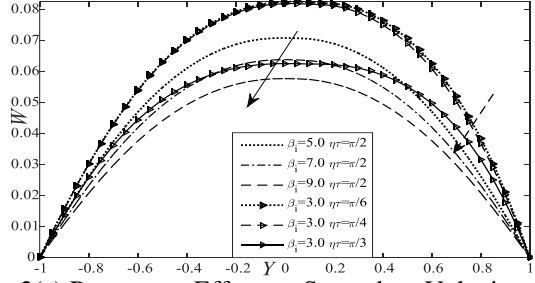
2(c) Parameters Effect on Secondary Velocity

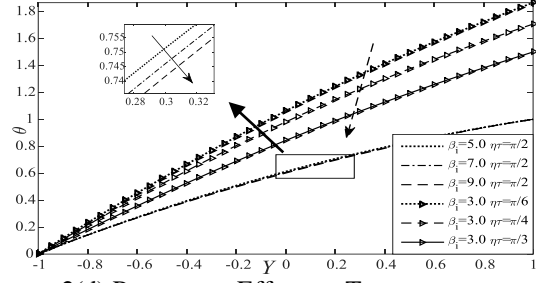
2(d) Parameters Effect on Temperature

**Figure 2.** Mesh Sensitivity test and Effects of Ion-slip parameter $(\beta_i)$ and phase angle $(\eta\tau)$; where, $\beta_e = 0.10$, $H_a = 3.00$, $R_e = 3.00$, $E_c = 0.10$, $P_r = 0.2$, $k_0 = 0.5$ and $\tau_D = 0.1$ at time $\tau = 2.00$ (Steady State)

**Mesh Sensitivity Test:** To obtain the appropriate mesh space for $m$ and $n$, the computations have been carried out for three different mesh spaces such as $(m,n) = (20, 20)$, $(40, 40)$, $(60, 60)$ as shown in Fig. 2(a). The curves are smooth for all mesh spaces and shows a negligible changes among the curves. Thus the mesh size $(m,n) = (40, 40)$ can be chosen as the appropriate mesh space.

Fig. 2(b,c,d) shows that the primary velocity increases with the increase of $(\beta_i)$ while the secondary velocity and temperature distributions decreases with the increase of $(\beta_i)$. Furthermore, the primary velocity, secondary velocity and temperature distributions decrease with the increase of $(\eta\tau)$.

**(Stationary Plate)** **(Oscillatory Plate)**

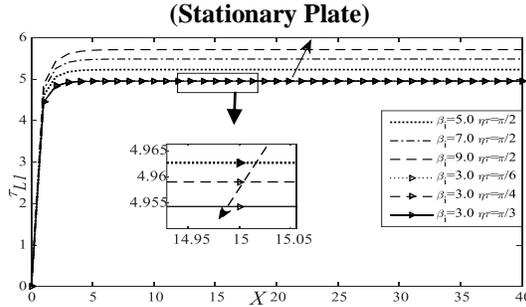
3(a) Parameters Effect on Local Shear Stress

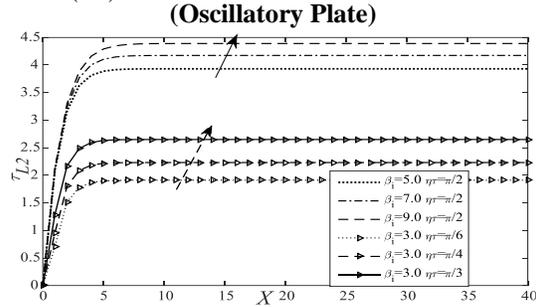
3(b) Parameters Effect on Local Shear Stress

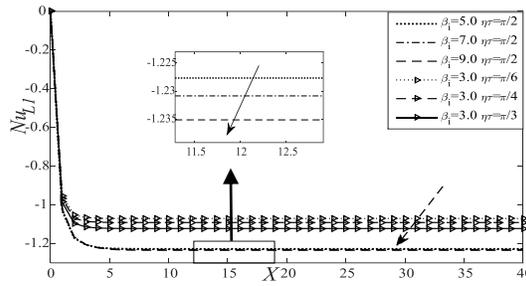
3(c) Parameters Effect on Local Nusselt Number

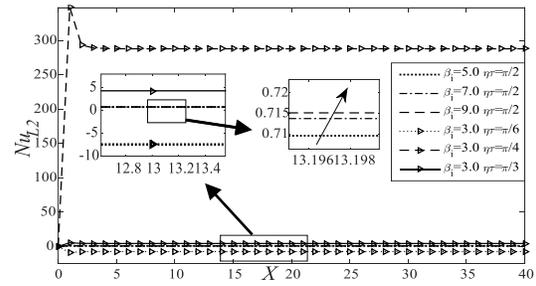
3(d) Parameters Effect on Local Nusselt Number

**Figure 3.** Effects of Ion-slip parameter $(\beta_i)$ and phase angle $(\eta\tau)$; where, $\beta_e = 0.10$, $H_a = 3.00$, $R_e = 3.00$, $E_c = 0.10$, $P_r = 0.2$, $k_0 = 0.5$ and $\tau_D = 0.1$ at time $\tau = 2.00$ (Steady State)

Fig. 3 shows that the local shear stress increases with the increase of $(\beta_i)$ while it decreases with the increase of $(\eta\tau)$ at stationary plate. At oscillatory plate, the local shear stress increases with the increase of $(\beta_i)$ and $(\eta\tau)$ both. From this figure, it is also seen that local Nusselt number decreases with the increase of $(\beta_i)$ and $(\eta\tau)$ both at stationary plate. At oscillatory plate, it increases with the increase of $(\beta_i)$ while it fluctuates with the increase of $(\eta\tau)$.

## CONCLUSION

The explicit finite difference solution has been carried out for the unsteady viscous incompressible Bingham fluid flow through oscillatory porous plate with Ion-Slip and Hall current. The numerical solution has found to be converged for $P_r \geq 0.08$, $\beta_i \geq 2$, $H_a \leq 20$, $k_0 \leq 8$ $(k \neq 2)$ and $R_e \geq -0.011$ with $\beta_e = 0.10$, $E_c = 0.10$, $\Delta Y = 0.05$ and $\Delta\tau = 0.0001$. At the steady-state time, the results are discussed graphically for different values of important parameters, such as Ion-slip parameter $(\beta_i)$, where $\alpha_e = 1 + \beta_i\beta_e$ and phase angle $(\eta\tau)$. For brevity, the effect of other parameters like Hall parameter $(\beta_e)$, Hartmann number $(H_a)$, Reynolds number $(R_e)$, Prandtl number $(P_r)$, Eckert number $(E_c)$, Permeability of porous medium $(k_0)$ and Bingham number $(\tau_D)$ are not shown. The specific findings of this investigation can be pointed out as follows:

1. The primary velocity increases with the increase of $(\beta_i)$ while it decreases with the increase of $(\eta\tau)$.
2. The secondary velocity and temperature distributions decrease with the increase of $(\beta_i)$ and $(\eta\tau)$ both.
3. The local shear stress increases with the increase of $(\beta_i)$ while it decreases with the increase of $(\eta\tau)$ at stationary plate. At oscillatory plate, the local shear stress increases with the increase of $(\beta_i)$ and $(\eta\tau)$ both.
4. The local Nusselt number decreases with the increase of $(\beta_i)$ and $(\eta\tau)$ both at stationary plate. At oscillatory plate, it increases with the increase of $(\beta_i)$ while it fluctuates with the increase of $(\eta\tau)$.

## REFERENCES

1. G. G. Stokes, "On the effect of the internal friction of fluids on the motion of pendulums", *Transactions Cambridge Philosophical Society*, (1851), Vol. **9**, pp. 8-106.
2. N. Datta, and R.N. Jana, "Oscillatory Magetohydrodynamic Flow past a Flat Plate with Hall Effects", *Journal of Physical Society of Japan*, (1975), Vol. **40**, pp. 1769-1474.
3. M. A. El-Hakiem, "MHD oscillatory flow on free convection–radiation through a porous medium with constant suction velocity", *Journal of Magnetism and Magnetic Materials*, (2000), Vol. **220**, No. 2-3, pp. 271-276.
4. O. A. Bég, J. Zueco, and H. S. Takhar, "Unsteady magnetohydrodynamic Hartmann–Couette flow and heat transfer in a Darcian channel with Hall current, ionslip, viscous and Joule heating effects: Network numerical solutions", *Communications in nonlinear science and numerical simulation*, (2009), Vol. **14**, Issue- 4, pp. 1082-1097.
5. A. Rao, S. Raju, and S. Sivaiah, "Finite element solution of heat and mass transfer in MHD flow of a viscous fluid past a vertical plate under oscillatory suction velocity", *Journal of Applied Fluid Mechanics*, (2012), Vol. **5,** Issue-3, pp. 1-10.
6. M. D. Hossain, M. A. Samad, and M. M. Alam, "MHD free convection and mass transfer flow through a vertical oscillatory porous plate with Hall, ion-slip currents and heat source in a rotating system", *Procedia Engineering*, (2015), Vol. **105**, pp. 56-63.
7. N. A. M. Zin, I. Khan, and S. Shafie, "The impact silver nanoparticles on MHD free convection flow of Jeffrey fluid over an oscillating vertical plate embedded in a porous medium", *Journal of Molecular Liquids*, (2016), Vol. **222**, pp. 138-150.
8. F. Sharmin, and M. M. Alam, "MHD Viscoelastic Fluid Flow along an Infinite Oscillating Porous Plate with Heat Source and Thermal Diffusion", *AMSE JOURNALS (AMSE IIETA publication-2017-Series: Modelling B)*, Vol. **86,** N°4, pp. 808-829.